# DNA Probabilities in *People v. Prince*: When are racial and ethnic statistics relevant?

## David H. Kaye[1]

*Arizona State University*

**Abstract:** When a defendant's DNA matches a sample found at a crime scene, how compelling is the match? To answer this question, DNA analysts typically use relative frequencies, random-match probabilities or likelihood ratios. They compute these quantities for the major racial or ethnic groups in the United States, supplying prosecutors with such mind-boggling figures as "one in nine hundred and fifty sextillion African Americans, one in one hundred and thirty septillion Caucasians, and one in nine hundred and thirty sextillion Hispanics." In *People v. Prince*, a California Court of Appeals rejected this practice on the theory that only the perpetrator's race is relevant to the crime; hence, it is impermissible to introduce statistics about other races. This paper critiques this reasoning. Relying on the concept of likelihood, it presents a logical justification for referring to a range of races and identifies some problems with the one-race-only rule. The paper also notes some ways to express the probative value of a DNA match quantitatively without referring to variations in DNA profile frequencies among races or ethnic groups.

## 1. Introduction

For two decades, the legal system has grappled with DNA evidence. The most difficult issues have not been technological, but statistical and logical. When a defendant's DNA matches that found at a crime scene, DNA analysts typically apply a population-genetics model to derive genotype frequencies from estimates of allele frequencies in the major racial or ethnic groups in the United States. But these estimates have been controversial. There have been objections to the quality and quantity of the data on DNA allele frequencies, the realism of the population-genetics models, and the manner in which the genotype frequencies are presented to judges and jurors.[2]

The case of *People v. Prince*[3] involves the problem of presentation rather than computation. In *Prince*, the California Court of Appeals stated that a genotype frequency estimate for a racial or ethnic group, no matter how well founded, is not relevant to determining whether the defendant is the source of the crime-scene DNA

---

[1] Arizona State University, Sandra Day O'Connor College of Law, McAllister & Orange Streets, P. O. Box 877906 Tempe, AZ 85287-7906, USA, e-mail: David.Kaye@asu.edu

[2] See generally 4 Modern Scientific Evidence: The Law and Science of Expert Testimony ch. 32 (David Faigman et al. eds., 3d ed. 2005-2006).

[3] 36 Cal.Rptr.3d 300 (Cal. Ct. App. 2005), rev. granted, 132 P.3d 210 (Cal. 2006), rev. dismissed, 142 P.2d 1184 (Cal. 2006).

unless there is "independent evidence" that the perpetrator is of the same race or ethnicity as the defendant. At the same time, the court held that sufficient evidence (other than the DNA match) that the defendant is the culprit can establish this preliminary fact and thus make the frequency in the defendant's racial or ethnic group relevant.

This chapter maintains that both these propositions are false. Relying on a likelihood-based theory of relevance, I first show that the population statistics for various "races" or ethnic groups ordinarily are relevant without regard to the defendant's race or ethnicity. They are relevant because they indicate that it is improbable that an innocent defendant would have a DNA genotype that just happens to match that of the actual perpetrator.

Second, I explain that, if it were true that frequencies in major subpopulations were only conditionally relevant, this condition normally could not be satisfied by other evidence that the defendant is guilty. Instead, evidence that makes it more probable that some *other* member of his racial or ethnic group (as opposed to any other such group) is the culprit would be necessary. Evidence that only singles out the defendant but otherwise provides no information about the race or ethnicity of the perpetrator does not help satisfy the court's condition.

After criticizing the reasoning in *People v. Prince* as to conditional relevance, I consider another objection to presenting the genotype frequency estimates for specific racial or ethnic groups. This is the possibility, emphasized in *Prince*, that a juror will jump to the conclusion that because the frequency in the defendant's racial and ethnic group is mentioned, the perpetrator must belong to this subpopulation. I argue that this concern, while appropriate in some cases, does not justify excluding reasonable estimates of genotype frequencies by racial or ethnic groupings.[4]

## 2. The court of appeals opinion

Patrick Paul Prince was charged with twelve counts of burglary, assault, and sexual crimes against five victims. In two of the attacks, "DNA matching his was found on a mask that each girl identified as having been worn by her attacker."[5] Comparable evidence was not available in the other crimes, but the prosecution maintained that the modus operandi was so distinctive that all had to have been committed by the same individual. A criminalist found DNA on the mask and discovered that it matched Prince's at nine STR loci. She testified about "a likelihood ratio that compared two different alternative possibilities, i.e., either the individual contributing the known reference sample contributed the evidence DNA and that is why the profiles matched; or the evidence DNA was contributed by some unknown, unrelated individual who happened to have the same DNA profile."[6] Using data from the FBI on three samples of about 200 Caucasians, 200 Hispanics, and 200 African-Americans, she concluded that "for the Caucasian population, the evidence DNA profile was approximately 1.9 trillion times more likely to match appellant's DNA profile if he was the contributor of that DNA rather than some unknown, unrelated individual; for the Hispanic population, it was 2.6 trillion times more likely; and for the African-American population; it was about 9.1 trillion times more likely."[7]

---

[4] The Supreme Court of California reached the same conclusion and also rejected the view that subgroup frequencies are only conditionally relevant in *People v. Wilson*, 136 P.3d 864 (Cal. 2006). The state supreme court also granted review in *Prince* and then dismissed the grant of review "[i]n light of our decision in *People v. Wilson* ...." 142 P.3d 1183 (Cal. 2006). This resolves a split between the different divisions of the California Court of Appeals. In light of the opinion in *Wilson* and the granting of review in *Prince* itself, the Court of Appeals' opinion in *Prince* no longer has precedential value in California. The case remains noteworthy, however, both because courts in other jurisdictions might find it attractive and because it is a stark example of how seemingly simple statistics can generate considerable confusion in the legal system.

[5] 36 Cal.Rptr.3d at 303.

[6] *Id.* at 310.

The jury returned a verdict of guilty on all counts. The trial court sentenced Prince to prison for well over 75 years and imposed monetary penalties. Prince appealed, primarily on the ground that "this statistical evidence was irrelevant because the prosecution failed to present substantial evidence to prove that the perpetrator was Caucasian, Hispanic, or African-American."[8] The California Court of Appeals agreed that independent evidence had to establish the racial or ethnic identity of the perpetrator for any statistics on the frequency of a DNA genotype in that racial or ethnic group to be relevant. However, it reasoned that because there was substantial non-DNA evidence linking Prince, a Caucasian, to the two crimes, the statistic for Caucasians was relevant and properly admitted. It regarded the admission of the statistics for Hispanics and African-Americans as erroneous, but harmless. Despite the perceived errors at trial, the Court of Appeals thus affirmed the convictions.[9]

This result is correct, but not for the reasons offered by the court. It is wrong to say, as the court does, that "[t]he probative value (hence, the relevancy) of a [DNA] profile's frequency in an ethnic population depends on proof that the perpetrator belongs to that ethnic group."[10] Moreover, the court's thought that the other evidence against Prince made the likelihood ratio of 1.9 trillion relevant is inconsistent with the logic behind introducing the likelihood ratio as a measure of probative value. To explain why, we must first clear away a bit of legal and logical underbrush.

## 3. The meaning of relevance

"'Relevant evidence' means evidence ... having any tendency in reason to prove or disprove any disputed fact that is of consequence to the determination of the action."[11] Consequently, to the extent that a DNA match tends to prove that the biological trace evidence that apparently originated from the perpetrator of the crime came from the defendant's cells, it is relevant. A DNA match has this tendency when the probability of observing the matching trace evidence is greater if the defendant is the source than if someone else is.[12] If $E$ stands for the observations of the DNA genotypes, and if $D$ represents the hypothesis that the defendant (D) is the source, while $I$ represents the hypothesis any other individual is the source, then the evidence $E$ is relevant to prove $D$ if and only if $P(E|D) > P(E|I)$. In

---

[7]*Id.*

[8]*Id.*

[9]More precisely, it affirmed the convictions on ten counts and reversed the convictions on another two counts. *Id.* at 327.

[10]*Id.* at 304.

[11]Cal. Evid. Code §210; cf. Fed. R. Evid. 401 ("'Relevant evidence' means evidence having any tendency to make the existence of any fact that is of consequence to the determination of the action more probable or less probable than it would be without the evidence.").

[12]See D.H. Kaye, The Relevance of "Matching" DNA: Is the Window Half Open or Half Shut?, 85 J. Crim. L. & Crimin. 676 (1995); Richard Lempert, Some Caveats Concerning DNA as Criminal Identification Evidence: With Thanks to the Reverend Bayes, 13 Cardozo L. Rev. 303 (1991).

other words, $E$ is relevant (and tends to prove $D$) if and only if

$$LR = \frac{P(E|D)}{P(E|I)} > 1. \tag{1}$$

But how do we know that the likelihood ratio $LR$ exceeds 1? This is where the frequencies in *Prince* come in. If the evidence is very likely to arise when D is the source and very unlikely when someone else is, then the numerator in (1) is much greater than the denominator, i.e., $LR \gg 1$. $E$ is then not only relevant but also highly probative.[13] Putting to the side such complications as the possibility of cross-contamination that would produce a false-positive match, if the matching DNA genotype is very rare among individuals other than D, then $LR \gg 1$. Thus, the figures quoted in *Prince* are relevant if they tend to show that condition (1) holds.

## 4. A hypothetical case

To establish that the DNA statistics in *Prince* tend to show how probative the DNA match is, it may be helpful to consider a hypothetical case. Suppose that a man wearing a mask murdered a student in her dormitory room at two a.m. The building always is locked from midnight until 6 a.m., and tapes from surveillance cameras on all possible entrances to the building demonstrate that no one entered or left the building in that period. The police arrived at 6 a.m., when they sealed and searched the building. They recovered the mask in the hallway and obtained a multilocus genotype from saliva extracted from the mask. They also took DNA samples from all people who were in the building overnight. One of these individuals (D), was found to have matching DNA ($M_D$). DNA from the remaining $n$ individuals was collected. In each instance, the DNA did not match that from the mask ($M_i^C$). Unless the unique match can be attributed to handling or laboratory error or fraud, the conclusion that D's DNA was on the mask is inescapable.[14] Although no formal analysis is necessary in this case, in more realistic situations and in understanding the implications of DNA testing of more than one individual, it is useful to specify the likelihoods.[15] Here, the DNA tests provide diagnostically valuable evidence with respect to $n+1$ hypotheses – one for D, and another $n$ corresponding to every other person in the building. One hypothesis is that D is the source of the DNA on the mask. As before, we denote this by $D$. The other hypotheses are that each individual I (other than D) is the source. These hypotheses can be denoted by $I_i$ (where $i = 1, \ldots, n$). With this notation, the evidence $E$ is the match to D and only D:

$$E = M_D \bigcap_{i=1}^{n} M_i^C. \tag{2}$$

---

[13]Arguments for using likelihood ratios to measure probative value are presented in D.H. Kaye & J.J. Koehler, The Misquantification of Probative Value, 27 Law & Hum. Behav. 645, 645–659 (2003).

[14]Being the source of the DNA on the mask is not logically equivalent to being the assailant. To draw that inference, we would need assure ourselves that the assailant wore this mask, that he left his DNA on it, and that this is the DNA that was detected.

[15]See generally David J. Balding, Weight-of-Evidence for Forensic DNA Profiles (2005); Peter Donnelly & Richard D. Friedman, DNA Database Searches and the Legal Consumption of Scientific Evidence, 97 Mich. L. Rev. 931 (1999).

If the hypothesis $D$ is true, then $P(E|D) > 0$. Indeed, assuming that the match is not the result of laboratory or handling error, then D is certain to match and everyone else is almost certain not to match (when D's genotype is very rare and hence very unlikely to be seen in the $n$ other individuals who were tested). In these circumstances, $P(E|D)$ is close to 1.

As for the denominator, if D is not the source of the DNA, but some other individual $I_a$ in the building is, then D is very unlikely to match (as are all the other innocent individuals) but the assailant $I_a$ is certain to match (again assuming that the reported genotypes are the true genotypes of each individual). Yet, every other possible suspect did not match. The evidence is thus impossible under the hypothesis that a specific person ($I_a$) other than D is the source. The conditional probability of $E$ is therefore

$$(3) \qquad P(E|I_a) = 0.$$

Hence, the likelihood ratio is a number close to 1 divided by 0, which is undefined (or infinite, as it must be if the claim that $E$ deductively implies $D$ is correct).

To shed light on the court's claim that statistics on the frequency of the DNA genotype in a given racial or ethnic group are irrelevant unless there is proof that the perpetrator is a member of this group, we introduce two factual variations. First, assume that D was the first person tested, that he matched, and that no one else was tested. This evidence is less complete but more like that in *Prince*. In contrast to (2), now

$$(4) \qquad E = M_D.$$

For this evidence, the numerator of (1) is simply the likelihood that $D$ would match if he were the source. This is $P(E|D) = P(M_D|D) = 1$. The denominator is the likelihood that $D$ would match if he were not the source of the DNA. Continuing to assume that there are no explanations such as fraud or error, and assuming that $D$ was not selected for testing based on his genotype (or anything correlated to it), he can be regarded as a random draw from the population. The probability that he would have the misfortune of sharing the actual assailant's genotype is simply the frequency of this genotype in the population. Denoting this frequency as $f$, we can write $P(E|I_a) = P(M_D|I_a) = f$. We conclude that the likelihood ratio is

$$(5) \qquad \frac{P(E|D)}{P(E|I_a)} = \frac{1}{f}.$$

Second, assume that $\frac{1}{2}$ of the dormitory's residents are Caucasian, $\frac{1}{8}$ are Hispanic, $\frac{1}{8}$ are African-American, and the remaining $\frac{1}{4}$ have other ancestries. Assume further that $f$ is not known but that the genotype frequencies within the first three groups are known to be $f_C$, $f_H$, and $f_A$, respectively. It would be better to learn $f$ itself, but $f_C$, $f_H$, and $f_A$ are useful in indicating the magnitude of $f$. The jury can make better use of the evidence of $M_D$ if it knows how rare the genotypes are in several major population groups (or even one such group) than if it has no frequency information and must resort to uninformed guessing. The figures can assist the jury in deciding what value to attach to the finding of a match to the defendant. The criminalist in *Prince* made this explicit by testifying to the likelihood ratio in (5).

Furthermore, even if one does not accept the likelihood ratio as the measure of probative value, it still should be clear that the three frequencies are at least relevant. They help the jury to gauge how surprising $M_D$ is under the hypothesis that the culprit is unrelated to D and that D just happens to have the same genotype.[16] In other words, they convey a sense of the *p*-value for the null hypothesis $I_a$. If these numbers are as infinitesimal as the criminalist asserted, then a juror can reject the hypothesis that the match is an unfortunate coincidence.

## 5. Ethnicity as a "preliminary fact"

The court of appeals rejected this seemingly ineluctable conclusion. It wrote that:

> [A]ny [frequencies] that do not [come from] the perpetrator's racial group are irrelevant, of themselves, to establish that the defendant is likely the perpetrator.... The selection of three individual ethnic databases, even assuming they represent the three largest population groups, is insufficient for this purpose because they have no value independent of the ethnicity of the perpetrator. All such evidence tells the jury is that the DNA profile is statistically rare in those population groups. It neither excludes nor includes the perpetrator as a member of any of those groups, nor does it specifically identify the defendant as being in the same population group as the perpetrator.[17]

This is not a claim that the frequency evidence is relevant but inadmissible because it will confuse the jury. It is not a claim that the population genetics model underlying the estimate is speculative or not generally accepted. It is not a claim that categories like "Caucasian," "Hispanic," and "African American" are socially constructed and devoid of essential meaning. It is not a claim that the frequencies fail to account for the probability of laboratory or handling error. The court does not bar the introduction of the racial and ethnic frequencies on any of these possible grounds. Rather, it asserts that the frequencies are logically irrelevant unless another fact first is established on the basis of other evidence: "The probative value (hence, the relevancy) of a profile's frequency in an ethnic population depends on proof that the perpetrator belongs to that ethnic group."[18]

In many cases, some evidence of the perpetrator's ethnicity will be available. If another dormitory resident had noticed a dark-skinned individual slipping out from the deceased's room, the court's reasoning would permit $f_A$ and $f_H$ to be admitted (but not $f_C$ (?) even though the ethnic identification is fallible and there is a broad range of skin coloration within all the subgroups). In a "general-population case,"[19] however, no evidence points to any subgroup member (racial, ethnic, or otherwise). The applicable frequency is that in the general population ($f$).

Yet, *Prince* insists that "because the frequency with which a genetic profile occurs varies according to ethnicity, it cannot be concluded the defendant likely left the biological evidence at the crime scene or on inculpatory matter absent some evidence that the person who left it belongs to the same ethnic group as the defendant."[20] Well, why not? Suppose that genotype frequencies were produced not only for the three subpopulations in *Prince* but for ten or twenty other subgroups – from Sephardic Jews, to Irish Catholics, to Naxi Dongba – and these frequencies were all quite small. There still would be no evidence that "the person who left [the trace DNA] belongs to the same ethnic group as the defendant." Such is the nature of a general-population case. Yet, it would be obtuse to insist that the comprehensive range of frequencies is not relevant in such a case because "the evidence does not support a finding of preliminary fact that the perpetrator belongs to the same racial/ethnic group."[21]

---

[16] See D.H. Kaye et al., The New Wigmore, A Treatise on Evidence: Expert Evidence §12.3.4 (2004).

[17] *Prince*, 36 Cal.Rptr.3d at 313–14.

[18] *Id.* at 304; see also *id.* at 308 ("the perpetrator's ethnicity should have been established independently").

[19] David H. Kaye, DNA Evidence: Probability, Population Genetics, and the Courts, 7 Harv. J. L. & Tech. 101 (1993). Of course, even in a "subpopulation case" in which there is information as to the perpetrator's race or ethnicity, recourse to other subgroups may be appropriate. The information pointing to a single racial or ethnic group is unlikely to be so conclusive as to preclude the jury from considering other possibilities.

[20] 36 Cal.Rptr.3d at 322.

In fact, the *Prince* court ultimately seems to reject its own demand for independent proof of the perpetrator's ethnicity. It expands "the perpetrator's population" to encompass "the general population" when it writes that "it is not enough to show that the genetic profile is rare in a certain number of ethnic populations. Instead, it must be shown either that the genetic profile is rare in the perpetrator's ethnic group *or in the general population*, as these are the perpetrator's populations."[22]

If the court is serious about allowing proof of the frequency "in the general population," then the puzzle in *Prince* is why proof that "the genetic profile is rare" in a range of ethnic populations does not show that it is rare in the general population. The court suggests that something is lacking – "a showing that [the frequencies] represent not just major population groups but the general population as a whole."[23] However, the opinion does not reveal what might constitute the requisite showing, although it alludes to the possibility of expert testimony.[24] Presumably, an expert might opine that inasmuch as the general population is composed of major population groups, the figures for these latter groups determine the figure for the population as a whole. But this is an algebraic truth rather than a substantive insight. If each major population group represents a proportion $p_j$ of the total population, with a genotype frequency $f_j$ in each group, then the frequency in the population is $f = \sum p_j f_j$. Because each $p_j$ is between 0 and 1, this weighted average must lie between the largest and the smallest $f_j$. Inasmuch as courts can take judicial notice of mathematical theorems, expert testimony is not even necessary when the "major population groups" exhaust the population.

In *Prince*, however, the criminalist did not give frequencies for a full partition of the general population. Only Caucasian, Hispanic, and African-American frequencies were presented. Even so, if these estimates – each in the trillionths – are credible, then the frequencies in the omitted groups would have to be enormously higher for the incriminating genotype to be anything other than "rare ... in the general population." The limited number of subpopulations described in *Prince* makes the inference from the three frequencies to the population something less than a mathematical truism, but evidence need not be conclusive to be relevant. It need only have "any tendency in reason to prove or disprove any disputed fact ...."[25] The three numbers surely had some tendency to prove that the DNA match was unlikely to be the result of coincidence. Indeed, as Table 1 shows, they arguably covered 94% of the population of Kern County, where the crimes occurred.[26]

---

[21] *Id.* at 305. Another objection to the court's rule is that "race/ethnicity" is not well defined. Certainly, society is not partitioned into distinct genetic groups, but the court's rule does not presuppose any scientifically objective reality. Rather, it states that there must be independent proof that the defendant falls into whichever category the prosecution adopts.

[22] *Id.* at 322.

[23] *Id.* at 323.

[24] *Id.* at 312.

[25] Cal. Evid. Code 210.

[26] The opinion does not specify the locations of the residences of the five victims. The DNA samples from the two victims were analyzed at the Kern County laboratory, and the trial was held in Kern County. Table 1 indicates that expanding the boundaries beyond Kern County would not alter materially the racial and ethnic population proportions.

TABLE 1
*Population percentages (2000)*

|          | Kern County | California | USA  |
|----------|-------------|------------|------|
| White    | 49.5        | 46.7       | 69.1 |
| Hispanic | 38.4        | 32.4       | 12.5 |
| Black    | 6.0         | 6.7        | 12.3 |
| Total    | 93.9        | 85.8       | 93.9 |

Source: U.S. Census Bureau, State and County Quick Facts, available at http://quickfacts.census.gov/qfd/states/06000.html & http://quickfacts.census.gov/qfd/states/06/06029.html.

TABLE 2
*Kern county 9-locus genotype frequency for possible frequencies in the groups not included in the criminalist's statistics*

|             | Percent of County $p_j$ | Genotype Frequency $f_j$ | Product $p_j f_j$ | Total $\sum p_j f_j$ |
|-------------|-------------------------|--------------------------|-------------------|----------------------|
| 1. White    | 49.5                    | $5.26 \times 10^{-10}$   | $2.61 \times 10^{-10}$ |                      |
| 2. Hispanic | 38.4                    | $3.85 \times 10^{-10}$   | $1.48 \times 10^{-10}$ |                      |
| 3. Black    | 6.0                     | $1.10 \times 10^{-10}$   | $6.59 \times 10^{-12}$ |                      |
| 4. Other    | 6.1                     | $f_1 = 5.26 \times 10^{-10}$ | $3.21 \times 10^{-11}$ | $4.47 \times 10^{-10}$ |
|             |                         | $10 f_1$                 | $3.21 \times 10^{-10}$ | $7.36 \times 10^{-10}$ |
|             |                         | $100 f_1$                | $3.21 \times 10^{-9}$  | $3.63 \times 10^{-9}$  |
|             |                         | $1000 f_1$               | $3.21 \times 10^{-8}$  | $3.25 \times 10^{-8}$  |

Using the percentages in Table 1 for Kern County, Table 2 shows the genotype frequency in the county for a range of values of the frequency in the 6% of the population not covered by the criminalist's statistics. Even if the nine-locus genotype frequency among the people not included in the three groups were a thousand times larger than the largest frequency in those groups, the frequency for the general country population would be only about 3 in 100,000,000.

Of course, one can be skeptical of the criminalist's likelihood ratios (which have been used to derive the frequencies $f_j$ in Table 2). But this does not render them *irrelevant* for the purpose of assessing how rare the nine-locus genotype is in the county, state, or national population. As with any other relevant evidence, the limitations in the statistics could have been explored on cross-examination.[27] Also, it could be argued that they are unfairly prejudicial or not generally accepted in the scientific community (the standard for admitting scientific evidence in California) and inadmissible for one of those reasons. But this is not what the *Prince* court said. The court contended that even if the frequencies in the three major groups are perfectly accurate, they are logically irrelevant and give no useful guidance to the jury. That position is untenable. The ethnicity of the perpetrator is not a "preliminary fact" that must be established by independent evidence.[28] In a general-population case, statistics on racial, ethnic, or geographic groups have some tendency to show how rare a DNA genotype is in the population of possible perpetrators. As such, they are relevant.

---

[27]Unless the failure of *Prince's* counsel to pursue this line of questioning amounted to constitutionally inadequate representation (an argument that the court rejected, 36 Cal.Rptr.3d at 326 n.33), it is not a ground for reversing a conviction.

[28]For further discussion, see D.H. Kaye, Logical Relevance: Problems with the Reference Population and DNA Mixtures in *People v. Pizarro*, 3 Law, Probability & Risk 211 (2004).

## 6. The misapplication of the preliminary-fact rule

Having incorrectly determined that the perpetrator's ethnicity is a preliminary fact that requires independent proof, the court in *Prince* proceeded to misapply this preliminary-fact rule. It reasoned that the preliminary fact was proved by evidence that linked the defendant and no one else in particular to the rapes:

> In the present case, there was no direct evidence of the perpetrator's ethnicity. However, direct evidence, such as a description from a percipient witness, is not the only means of establishing the preliminary fact. Instead, the requisite fact can also be established through other independent evidence (evidence not dependent upon the profile match, match frequency, or the defendant's ethnicity per se) that the defendant is the perpetrator. The logic is as follows: If independent evidence establishes that the defendant more likely than not is the perpetrator, and the defendant is Caucasian, then independent evidence establishes . . . that the perpetrator more likely than not is Caucasian. The preliminary fact of the perpetrator's ethnicity is thus sufficiently established so that match frequency statistics, computed from a Caucasian database, are relevant to prove the defendant's identity as the perpetrator.[29]

The "independent evidence" that the court deemed sufficient to establish the preliminary fact was

> the actual evidence . . . that defendant is the perpetrator. . . . [E]vidence presented by means of victim testimony showed that the perpetrator wore a ski mask and used a flashlight with a colored lens. Appellant possessed such items; therefore, the evidence establishes that appellant could be the perpetrator.[30]

And so, the court concluded that:

> [s]ince the record sufficiently establishes that appellant is the perpetrator and therefore shares the perpetrator's race, and appellant is Caucasian, the profile frequency statistics derived from the Caucasian database, as testified to by [the criminalist], were relevant. The evidence with respect to the other ethnic databases was not; however, its admission did not prejudice appellant.[31]

This reasoning could be applied in virtually every case in which some evidence beyond the DNA match incriminates the defendant. Such "independent evidence" need not be extensive. It need only "be sufficient evidence to enable a reasonable jury to conclude that it is more probable that the fact exists than that it does not."[32] As a result, the practical impact of the *Prince* rule would be minimal. As long as there is some non-DNA evidence to indicate that it is probable that defendant committed the alleged crime, the prosecution can introduce the statistic for the defendant's ethnic or racial group. If the prosecution errs by presenting other frequencies, the error will be harmless, for multilocus STR types are rare in all ethnic and racial groups.

Not only does the court's use of the non-DNA evidence against Prince (and only Prince) to justify admitting the statistic for Caucasians eviscerate its preliminary-fact requirement, but it also ignores the proper purpose for admitting the statistic. The only reason that DNA statistics are used is to assess *alternative* hypotheses by quantifying the probability of that the match would occur when an unrelated individual is the source of the trace evidence and this person just happens to have the same DNA genotype as the defendant. This unrelated individual might be of the same racial or ethnic group as the defendant, or he might not be. Evidence tending to show that the defendant D, and no other individual I, is responsible for the crime does nothing to clarify I's status. It bears no relationship to the denominator of the likelihood ratio in (5). It makes it more probable that the defendant is guilty, of course, but it does not affect the probability that if the defendant is not the perpetrator, someone of the same ethnicity is.

---

[29] 36 Cal.Rptr.3d at 325 (citations and footnote omitted).

[30] *Id.* at 326. The court also alluded to "evidence [set out] at length in the unpublished portion of our opinion . . . . Suffice it to say that items seized from appellant's residence and car, as well as appellant's reaction when asked to give a buccal swab, are pertinent to our analysis. Although none is sufficient standing alone, when considered cumulatively, they meet the . . . standard." *Id.*

[31] *Id.*

[32] *Id.* at 325.

Consequently, the evidence described in *Prince* is not the type of "independent evidence" that the court should demand. If foundational evidence were required, it would have to point not just to the defendant, but more broadly to other members his ethnic or racial group in contrast to other groups. Unless it can be said that Caucasians as a group are especially likely to posses ski masks and flashlights with a colored lenses, this evidence does not imply that the perpetrator is more probably Caucasian than anything else. As a result, it does not, on the court's premises, justify upholding the admission of the Caucasian genotype frequency.

## 7. The real concern: prejudice

Thus far, I have argued that in a general-population case, DNA genotype frequencies in major racial or ethnic groups are relevant. The *Prince* opinion denies this, but, at bottom, the court's concern could be prejudice, not relevance. In an earlier case, *People v. Pizarro*,[33] the same court of appeals introduced the flawed preliminary-fact analysis to correct the fallacious reasoning of the prosecution in that case that just because the defendant has a certain ancestry or ethnicity, whoever left the DNA has the same ancestry or ethnicity. The *Pizarro* court wrote that

> If the only way you can conclude the perpetrator fits a racial/ethnic category is to assume the perpetrator was the same race/ethnic background as the suspect then the reasoning is circular, i.e.: proof of the racial/ethnic background of the perpetrator depends on the racial/ethnic background of the suspect from which we infer a statistical probability that the perpetrator is the suspect. ... [W]e see no relevancy to a database selected because of the racial/ethnic background of the suspect/defendant. The problems created by employing assumed relevancy of the database are insidious. A jury hears an astronomical figure that not uncommonly depends for its relevance upon the very issue that they have to decide: is the defendant the perpetrator?[34]

In *Prince*, the court[35] expanded on this criticism of the "circular" and "insidious" prosecutorial reasoning.[36] It complained that

> Because of the way the evidence is presented, ... use of the defendant's ethnic population without reference to that of the perpetrator, as occurred in *Pizarro*, ineluctably points the finger at the defendant, even though it does not truly prove anything. On the other hand, it presents jurors with a very subtle, but very insidious, form of racial profiling: the assumption that the perpetrator and the defendant are the same race and, therefore, that the defendant is the perpetrator.[37]

---

[33]12 Cal.Rptr.2d 436 (Ct. App. 1992) (reversing and remanding); 3 Cal.Rptr.3d 21(Ct. App. 2003) (reversing again following the remand), *review denied* (Oct 15, 2003).

[34]12 Cal.Rptr.2d at 94 (footnote omitted).

[35]Judge Ardaiz wrote both opinions.

[36]The assumption that the defendant's race or ethnicity determines the relevant population is not confined to prosecutors. Defendants have used it to argue the prosecution's statistics are irrelevant or inapposite because they do not match the defendant's subgroup. E.g., People v. Mohit, 579 N.Y.S.2d 990 (Westchester Co. Ct. 1992) (reasoning that inbreeding among Shiite Muslims around Shushtar, Iran, was "of particular importance" because the defendant was a physician born in Shushtar; however, the alleged crime was a sexual assault of a patient in New York, not in Iran).

[37]36 Cal.Rptr.3d at 313.

Although this criticism is overstated,[38] the concern that the jury will misuse valid information can be grounds for its exclusion. In *Prince*, the court of appeals dismissed the state's argument that the frequencies in *Prince* were indicative of the general-population frequency, in part because it thought that "the statistical calculations did not provide such evidence," and in part because "they [were not] proffered *for that purpose*."[39] We already have seen that it is implausible to argue that frequencies in the trillionths in the three groups that collectively encompass some 94% of the county's population have no bearing on the question of whether the genotype is rare in the immediate area, but it is true that the criminalist did not make this point in so many words. Yet, for what other purpose could the three numbers have been offered? To trick the jurors into thinking that merely because the defendant was Caucasian, so was the culprit? Then why give the figures for Hispanics and African-Americans? Why tell the jury that this procedure is used regardless of the defendant's specific race or ethnicity? And why inform the jury that the point of the three statistics was to "extrapolate[] to the population at large"?[40]

Moving beyond the details of *Prince*, it seems that in a general-population case, the mere fact that a forensic scientist reports that a DNA genotype is rare in a variety of subpopulations is not prejudicial. If anything, presenting a spectrum of values should dissuade the jury from thinking that the suspect population is limited to people exactly like the defendant. If so, the concern that the jury will misuse the evidence in this way is not an adequate reason to exclude it.

Other procedures to avoid the risk of this cognitive error also could be employed, and the *Prince* court enumerates several of them:

> [W]hen the perpetrator's ethnicity is unknown, the most appropriate solutions would appear to be (1) to present the one most conservative frequency, without mention of ethnicity, or (2) assuming this method is scientifically valid and results in a frequency that is considered conservative, to present a single frequency calculation based on a general, nonethnic population database.... When frequency calculations that do not reference ethnicity are employed, the profile frequency evidence no longer tells jurors that if the defendant and the perpetrator share ethnicity, the likelihood the defendant is the perpetrator is some number. Instead, the evidence tells jurors that regardless of the perpetrator's ethnicity, the likelihood the defendant is the perpetrator is some number. Jurors then have to decide whether, in their minds, the genetic profile is sufficiently rare so as to be persuasive as to identification of the defendant as the

---

[38]The rhetorical flourish of "racial profiling" is a misnomer. In neither *Pizarro* nor *Prince* was the defendant selected for investigation or prosecution because of his race – subtly or otherwise.

[39]36 Cal.Rptr.3d at 317 (emphasis added).

[40]*Id.* at 309 (footnote omitted). These words appear in the context of the following explanation:

> What I need to do is determine how common is this profile in the population .... [¶] The ... three populations that I generally look at are Caucasian, Hispanic, African-American. And I have some statistic[al] software to help out with the calculation part. I enter the data in and the [software] will do the calculations for me and determine just how common the profile is, you know, in this reference population, which extrapolates to the population at large.

*Id.* The court offered the following interpretation of these remarks:

> [W]e read this as meaning that the frequency calculated based upon the sample (reference) Caucasian population extrapolates to the Caucasian population at large, the frequency calculated based upon the sample Hispanic population extrapolates to the Hispanic population at large, and the frequency calculated based upon the sample African-American population extrapolates to the African-American population at large.

perpetrator.[41]

The court fails to explain why it calls for the most "conservative" figure as opposed to the most accurate. Normally, the law should seek unbiased estimates. This is true even though the burden of persuasion in a criminal case is proof beyond a reasonable doubt. The beyond-a-reasonable-doubt standard applies to the totality of the evidence, not to each scrap of evidence.

As a practical matter, however, an expert or a party might prefer to make conservative assumptions in some situations, and the law should accept this choice. For instance, one party might wish to forestall objections about missing data in a statistical study by assuming that all the missing data would be adverse to its position. If the overall result still favors this party, then the other side is in no position to argue that the study should be excluded because some data were missing. With regard to DNA evidence, various methods for arriving at upper-bounds on genotype frequencies in virtually any subpopulation have been devised. The notorious "ceiling principle" is one.[42] More recently, a working group of a national commission observed that:

> The differences in [DNA genotypes] are mainly between individuals rather than between group averages. This means that the necessity for group classification could be avoided by using an overall U.S. database and an appropriately increased value of $\theta$ [a parameter that measures population structure]. This has been advocated by some to avoid invoking any group identification. A $\theta$ value of 0.03 would usually be appropriate.[43]

This would be one way to implement the *Prince* court's desire for a race-blind figure in a general-population case.[44]

## 8. Conclusion

In *People v. Prince*, the California Court of Appeals misapplied the legal doctrine of conditional relevance to conclude that where the defendant was Caucasian, it was error to introduce statistics on the frequency of a certain DNA genotype among Hispanics and African Americans. In a general-population case like *Prince*, DNA genotype frequencies in a range of racial or ethnic groups are relevant because they bear on the alternative hypothesis that someone in the general population

---

[41] *Id.* at 313-314 (citation omitted).

[42] See generally Committee on DNA Technology in Forensic Science: An Update, National Research Council, The Evaluation of Forensic DNA Evidence (1996) (concluding that the ceiling principle is not needed); Committee on DNA Technology in Forensic Science, National Research Council Committee on DNA Technology in Forensic Science, DNA Technology in Forensic Science (1992) (proposing the ceiling principle as a compromise); Eric S. Lander & Bruce Budowle, Commentary: DNA Fingerprinting Dispute Laid to Rest, 371 Nature 735 (1994) (defending the ceiling principle).

[43] National Commission on the Future of DNA Evidence, The Future of Forensic DNA Testing: Predictions of the Research and Development Working Group 5 (2000).

[44] In a subpopulation case – one in which the alternative hypothesis is that the perpetrator is from a specific racial or ethnic group – and "there is uncertainty about the population substructure, as with isolated tribes or communities, or possible unsuspected relatives," the working group proposed "the Sib Method" for producing an upper bound:

> The conditional match probability for a pair of sibs is determined mainly by simple Mendelian rules and is relatively unaffected by allele frequencies (which may differ among population subgroups) and unsuspected substructure, inbreeding, or presence of relatives. Since no other relatives are as close as sibs, the match probability for sibs provides a rough upper limit for the actual match probability.

*Id.* at 4.

other than the defendant is the source of the DNA sample that incriminates the defendant. When introduced for this purpose (and it is hard to see what other purpose they could be introduced for), proof of these frequencies should not be excluded as irrelevant.

At the same time, there have been cases in which judges or juries have assumed that just because a defendant is a member of a particular race or ethnic group, the DNA genotype frequency within that group is the only relevant figure. This reasoning is potentially prejudicial to the defendant, and the *Prince* court was justified in discussing ways to avoid this cognitive error. Certainly, the prosecution should not overstate its case, but when the race or ethnicity of the perpetrator is uncertain, explaining that DNA genotype frequencies vary among racial groups and accurately displaying this variation does not exaggerate the evidence.

**Acknowledgments.** This paper benefited from comments by David Balding, James Crow, George Sensabaugh, and Terry Speed. I am also very grateful to David Freedman for perspicuity, persistence, and patience in working with me over the years on efforts to explain statistical reasoning to lawyers and judges. His contributions are apparent and abundant in our chapters on statistics in two editions of the Federal Judicial Center's Reference Manual on Scientific Evidence and five editions of Modern Science Evidence: The Law and Science of Expert Testimony.